\documentclass{article}

\input epsf

\begin{document}
\setlength{\unitlength}{1mm}

\begin{titlepage}

\begin{flushright}
LAPTH 789-00\\
LPT-Orsay 00-37\\
April 2000
\end{flushright}
\vspace{1.cm}

\begin{center}
\large\bf
{\LARGE\bf An automatized algorithm to compute infrared 
           divergent multi-loop integrals}\\[2cm]
\rm
{T.~Binoth$^{a}$ and G.~Heinrich$^{b}$ }\\[.5cm]

{\em $^{a}$Laboratoire d'Annecy-Le-Vieux de Physique 
 Th\'eorique\footnote{UMR 5108 du CNRS, associ\'ee \`a 
              l'Universit\'e de Savoie.} LAPTH,}\\
      {\em Chemin de Bellevue, B.P. 110, F-74941 
           Annecy-le-Vieux, France}
        
        \medskip   
            
{\em $^{b}$Laboratoire de Physique Th\'eorique\footnote{UMR 8627 du
           CNRS} LPT,\\ 
           Universit\'e de Paris XI, B\^atiment 210,\\
           F-91405 Orsay, France} \\[3.cm]

\end{center}
\normalsize

%\begin{center}
%\LARGE{PRELIMINARY VERSION: 31/03/00.}\\
%\end{center}

\begin{abstract}
We describe a constructive procedure to separate overlapping
infrared divergences in multi--loop integrals. 
Working with a parametric representation in $D=4-2\epsilon$
dimensions, adequate subtractions lead to a Laurent series
in $\epsilon$, where the coefficients of the
pole-- and finite terms are sums of regular parameter integrals
which can be evaluated numerically.
We fully automatized this algorithm by implementing 
it into algebraic manipulation programs
and applied it to calculate numerically some  nontrivial 2-loop
 4-point and 3-loop 3-point  Feynman diagrams.
Finally, we discuss the applicability of our method
to phenomenologically relevant multi--loop calculations such as 
the NNLO QCD corrections for $e^+e^-\to$ 3 jets. 
\end{abstract}

\vspace{3cm}

\end{titlepage}

\section{Introduction}

The increasing experimental precision at present and future colliders 
requires fast progress in the calculation of higher order
corrections from the theoretical side. A crucial role is thereby 
played by the calculation of multi-loop integrals, which becomes
an increasingly challenging task as the number of loops and the number 
of kinematic invariants gets larger. 

In the case of two-point functions, up to four loop orders could be 
evaluated~\cite{surguladze,larin}
by exploiting the integration-by-parts method~\cite{tkachov}
and recurrence relations. This powerful technique also has been 
generalized to calculate three-loop propagator-type diagrams 
in heavy quark effective theory~\cite{grozin}. 
Similarly, the presence of only one kinematic invariant allowed for the 
calculation of two-loop three-point functions
~\cite{gonsalves,lampe,matsuura,ussy,davydychevosland},
which could be applied to compute the partonic cross sections for DIS and 
 the Drell-Yan process to NNLO~\cite{zilvn, hamberg}. 
Physical processes with a more difficult kinematic
structure could not be computed to two-loop order yet, amplitudes 
exist only for very special cases~\cite{bdkallglu}.
A major reason is that two-loop integrals with four external
legs, involving more than one kinematic variable, constitute a more 
complex problem. 
Only very recently, 
the analytic calculation of the massless planar~\cite{smirnov} and 
non-planar~\cite{tausk} double box could be achieved by using Mellin-Barnes 
integration techniques. Furthermore, results for single-box
integrals with self-energy-- and vertex insertions have been given 
in~\cite{oleari} and techniques based on differential equations to 
reduce two-loop four point functions to a set of master integrals 
have been derived in~\cite{smirnoveretin,gehrmann}. The tensor reduction of massless  
double box integrals has been completed
in~\cite{Anastasiou}, where the second master integral needed in the 
reduction of crossed two-loop boxes is given. 
However, it is not clear at the moment whether the techniques developed 
so far are sufficient to solve more complicated 
higher order problems as well.

\smallskip

Hence, to obtain results for integrals which are beyond the scope of analytic 
integration methods, but also as a simple check of involved analytic 
calculations,  
it is desirable to have a method at hand which allows the calculation of  
multi-loop integrals numerically. 
For two-loop box integrals with internal masses, thus with no infrared divergence,  
semi-numerical approaches have been designed in
~\cite{kreimer}. 
The massless case requires a different approach because of
the presence of infrared divergences, which first have to be 
extracted from the integral in order to make it amenable to 
numerical evaluation.
 
\smallskip
In perturbative QCD, the calculation of infrared safe quantities 
has to be organized such that the infrared poles stemming from 
virtual and real higher order corrections cancel. 
At next-to-leading order usually semi-analytical methods are used to 
achieve this cancelation. 
As an alternative, a completely numerical method 
has been developed in~\cite{soper}.
The algorithm allows for the  summation of  
the contributions from different  
cuts of a graph  {\em before} integration, such that 
the  cancelation of soft and collinear divergences is built in as 
imposed by  unitarity. 
Nevertheless, since the method requires the deformation of the
multi-dimensional integration contours to avoid fake singularities,  
its general applicability to NNLO calculations is highly nontrivial. 

\smallskip

Hence a general {\em local} subtraction procedure to separate infrared 
divergences from  {\em individual}  graphs of arbitrary loop order would be 
very useful. 
The IR singularity structure of higher-loop Feynman diagrams
 was investigated in \cite{Kinoshita_papers} in four dimensions, and
later in~\cite{colsopster} 
in the context of dimensional regularization and factorization in QCD. 
Working in dimensional regularization,
subtraction procedures are well known for  
UV poles, and also for
IR divergences present in Euclidean space (see e.g. \cite{smirnov:buch}).
On the other hand, no general subtraction scheme 
for soft and collinear IR singularities arising in Minkowski space is known
for individual graphs.
The method we present in this paper has been designed to isolate  
poles in the dimensional regulator $\epsilon$ 
for an arbitrary Feynman graph. 
Although the method  also works for one-loop integrals,  
its virtues show up rather in
two-- or higher loop integrals with $N\ge 3$ external legs, at least one 
of them being massless. 
It allows one to disentangle overlapping soft and collinear 
divergent regions in Feynman parameter space by dividing the latter into
sectors where  parameters can get singular only in an independent
manner. Then, by adding and subtracting adequate counterterms, one can
isolate the singular parts and perform the integrations over the 
corresponding parameters analytically. The remaining regular integrals 
are in general too complex for analytical integration, but they can
be integrated numerically. 
This procedure is quite 
general and can in principle be applied to graphs with an arbitrary   
number of loops and legs, 
the limitations being only disk space and computing time. 

The algorithm can be divided into four blocks. In the first one, 
the $\delta$-distribution constraint on the Feynman parameters is 
eliminated in a particular way and in the second one, 
the singular contributions are isolated in parameter space. 
The third block consists of a subtraction procedure for the
$1/\epsilon^m$ poles, producing a  set $C_m$ of finite functions of the 
Feynman parameters as coefficients of each order--$m$  pole.
In the fourth block, the integrations over the 
Feynman parameters in these functions are performed. 
The coefficient functions of the leading and subleading pole of a graph
are in general simple enough to be integrated analytically with an 
algebraic manipulation program, whereas the remaining functions
have to be integrated numerically.

The paper is organized as follows. In section two, we outline the
algorithm. 
In section three, 
some examples are given in order to show the applicability of the method
to quite different types of integrals. First we treat some 
two-loop four-point functions with one external leg off-shell 
as an example for a three-scale problem
which could not be solved analytically yet. Then three-loop 
three-point graphs with two external legs on-shell and 
eight and nine propagators, respectively, are calculated.  
Section four contains a discussion of available analytical/numerical 
results for elements entering the calculation of various processes
of phenomenological relevance, pointing out where our method  
improves the situation.    

\section{The algorithm}

In this section we describe the algorithm to treat a general
$D$--dimensional scalar $L$--loop
Feynman diagram with $N$ propagators.
If $E$ is the number of external legs with momenta $\{p_1,\dots p_E\}$,
$L$ the number of loop momenta $\{k_1,\dots ,k_L\}$, $\{m_1,\dots m_N\}$ 
the (not necessarily nonzero) masses of the propagators 
$P_{j\in \{1,\dots ,N\}}$,
and $d{\cal K}_{m\in \{1,\dots ,L\}}=d^Dk_m/[i\pi^{(D/2)} ]$, a general
scalar Feynman diagram $G$ can be written as
\begin{eqnarray}
G &=& \int d{\cal K}_1\dots d{\cal K}_L \prod\limits_{j=1}^{N} P_{j}(\{k\},\{p\},m_j^2)
\end{eqnarray}  
In the present paper we will not deal with powers of propagators different from one.
Nevertheless, higher (and even non-integer) powers of propagators can be treated
with basically the same algorithm.   
Introducing Feynman parameters, the integral 
can be expressed in terms of a 
symmetric $(L\times L)$--matrix $M$,
an $L$-vector $Q$ (with 4-vectors in each component) and a scalar function $J$.
The contraction of Lorentz indices is indicated by a dot.
\begin{eqnarray}\label{EQ_mixed_rep}
G =  \Gamma(N)\int d^N x \,\, \delta(1-\sum_{l=1}^N x_l)
\int d{\cal K}_1\dots d{\cal K}_L 
\left[ 
       \sum\limits_{j,l=1}^{L} k_j\cdot k_l \, M_{jl} - 
       2\sum\limits_{j=1}^{L} k_j\cdot Q_j +J 
                                              \right]^{-N}
\end{eqnarray}  
After having shifted the loop momenta to get rid of the linear
term, one obtains after momentum integration 
the following parameter representation of the graph $G$.
\begin{eqnarray}\label{EQ:param_rep}
G &=& (-1)^N\Gamma(N-LD/2)\int
\limits_{0}^{\infty} d^Nx\, 
\delta(1-\sum_{l=1}^N x_l)\,
\frac{{\cal U}^{N-(L+1) D/2}}{{\cal F}^{N-L D/2}}\\
 &&\nonumber\\
\mbox{where} \qquad \quad &&\nonumber\\
{\cal F}(\vec x) &=& \det (M) 
\left[ J- \sum\limits_{j,l=1}^{L} Q_j\cdot Q_l \, M^{-1}_{jl}
 \right]\label{DEF:F}\\
{\cal U}(\vec x) &=& \det (M) 
\end{eqnarray}  
As is well known, the parametric
representation is determined by two functions which we call 
${\cal U}$ and ${\cal F}$. They are defined by
the topology of the corresponding 
Feynman diagram \cite{Nakanishi,zavialov,itzyksonzuber}. 
${\cal F}$ contains the Mandelstam
variables related to the different cuts of the graph.

A necessary condition
for the presence of infrared divergences is that the Landau 
equations~\cite{landau,itzyksonzuber} are fulfilled. 
A representation of the Landau equations 
following  from Eq.~(\ref{EQ_mixed_rep}) is given by
\begin{eqnarray} \label{EQ:Landau_eq}
k_l^{\mu} &=& \sum\limits_{j=1}^{L} M^{-1}_{lj} Q^{\mu}_j \nonumber\\
{\cal F} &=& 0
\end{eqnarray}
The function ${\cal U}$ cannot lead to infrared divergences of the 
graph, since giving a mass to all external legs would not change ${\cal U}$.
Apart from the fact that the graph may have an overall UV divergence contained in
the overall $\Gamma$-function (see Eq.~(\ref{EQ:param_rep})), UV subdivergences
may also be present. A necessary condition for these is the vanishing 
of ${\cal U}$. In this way
the UV poles can be identified and treated according to standard
renormalization procedures.
Infrared poles in $1/\epsilon$ come from the parameter region where
some Feynman parameters are small such that ${\cal F}$ vanishes.
The Feynman parameter integrations which lead to poles can be related to
a kinematical configuration in momentum space by using 
Eqs.~(\ref{EQ:Landau_eq}).
The IR poles come from soft and/or collinear momentum configurations 
where propagators do not describe  virtual particles anymore but rather 
the propagation of an on-shell particle together with
eventual splittings\footnote{This can be visualized by defining the
{\it reduced} graph~\cite{sterman} of a diagram, which is the diagrammatic
representation of solutions of the Landau equations.}. 
The soft/collinear
poles of multi--loop graphs are the result of such kinematical 
situations in loop momentum space. These singular regions 
are generally not separated from each other in momentum 
(or equivalently in Feynman parameter space), they are overlapping.   

Our aim is to disentangle these regions 
of  overlapping IR divergences. To this end we use a method called
sector decomposition\footnote{This method
was of some importance in the history of UV regularization, i.e. to  
establish the BPHZ method. It was used by 
Hepp~\cite{hepp} to deal with overlapping UV divergences.}. 
Iterated application will lead to 
a set of integrals in which the infrared singular behaviour is not contained
in complicated functions anymore, but in simple products of Feynman parameters
raised to some power, times  remnants of the functions ${\cal F}$, ${\cal U}$,
whose structure is  such that they neither lead to a pole anymore, nor
change the exponent of the respective poles.
In more detail, the procedure consists of four basic building blocks:

\subsection*{Part I \quad Generation of primary sectors}

In the first part of the algorithm we split the integration domain into
$N$ parts and eliminate the $\delta$--distribution in such a way that the remaining 
integrations are from 0 to 1.  
To this end we decompose the integration range as follows
\begin{eqnarray}
\int_0^{\infty}d^N x =
\int_0^{\infty}d^N x  \prod\limits_{j=1}^{N}\theta(x_j\ge 0) = 
\sum\limits_{l=1}^{N} \int_0^{\infty}d^N x
\prod\limits_{\stackrel{j=1}{j\ne l}}^{N}\theta(x_l\ge x_j\ge 0)
\end{eqnarray} 
where the $\theta$-function is defined as
\begin{displaymath}
\theta(x \ge y)=\left\{\begin{array}{ll}
              1& \mbox{if } x\ge y \mbox{ is true}\\
              0& \mbox{otherwise}\end{array}
              \right.
\end{displaymath}
The integral is now split into $N$ domains corresponding   
to $N$ integrals $G_l$ from which we extract a common factor:
 $G=(-1)^N \Gamma(N-LD/2) \sum_{l=1}^{N} G_l$. In the  integrals $G_l$
we integrate out $x_l$ by using the $\delta$--distribution after having 
done the substitution 
\begin{eqnarray}
x_j = \left\{ \begin{array}{lll} x_l t_j     & , & j<l \\
                                   x_l         & , & j=l \\
                                   x_l t_{j-1} & , & j>l \end{array}\right.
\end{eqnarray} 
Because of homogeneity, $x_l$ factorizes completely in the functions
${\cal U}(\vec x) \rightarrow {\cal U}_l(\vec t\,)\, x_l^L$ and
${\cal F}(\vec x) \rightarrow {\cal F}_l(\vec t\,)\, x_l^{L+1}$
and thus, using $\int dx_l/x_l\,\delta(1-x_l(1+\sum_{k=1}^{N-1}t_k ))=1$, one obtains 
\begin{eqnarray}\label{EQ:primary_sectors}
 G_l &=& \int\limits_{0}^{1} d^{N-1}t 
\frac{ {\cal U}_l^{N-(L+1)D/2}}{ {\cal F}_l^{N-L D/2}} \quad , \quad l=1,\dots N 
\end{eqnarray} 
Note that the singular behaviour leading to $\epsilon$--poles
still comes  from the regions of small $t$'s. This feature would be lost 
 if one integrated out the $\delta$--distribution
in a naive way, since this would produce poles at  upper 
limits of the parameter integral as well. 
The generated sectors
will be called {\em primary} sectors in the following.
The functions ${\cal U}_l$ and ${\cal F}_l$ are polynomials
in the parameters $t_j$. 
 
\subsection*{Part II \quad Iterated sector decomposition}

The second part of the algorithm consists 
of the iterated application of sector decomposition 
and a remapping of parameter space to the unit cube in order to 
disentangle the overlapping
singular regions of the integrands. 
Starting with Eq.~(\ref{EQ:primary_sectors}) one repeats the following 
steps until complete separation of overlapping regions is achieved.
\begin{description}
\item[II.1:] Determine a minimal set of parameters, say 
${\cal S}=\{t_{\alpha_1},\dots ,t_{\alpha_r}\}$, such that  
${\cal U}_l$, respectively  ${\cal F}_l$, vanish 
if the parameters of ${\cal S}$ are set to zero. ${\cal S}$ is generally
not unique. Additional selection criteria can be introduced 
to choose an ${\cal S}$ which does not lead to a large 
number of subsequent sector decompositions.
\item[II.2:] Decompose the corresponding $r$-cube into $r$ {\em subsectors}.
\begin{eqnarray}
\prod\limits_{j=1}^r \theta(1\geq t_{\alpha_j}\geq  0)=
\sum\limits_{k=1}^r \prod\limits_{\stackrel{j=1}{j\ne k}}^r 
\theta(t_{\alpha_k}\geq t_{\alpha_j}\geq 0)
\end{eqnarray}
 \item[II.3:] Remap  the variables to the unit cube in each new 
 subsector by substituting
 \begin{eqnarray}
t_{\alpha_j} \rightarrow 
\left\{ \begin{array}{lll} t_{\alpha_k} t_{\alpha_j} &,&j\not =k \\
                           t_{\alpha_k}              &,& j=k  \end{array}\right.
\end{eqnarray}
This gives a Jacobian factor of $t_{\alpha_k}^{r-1}$. By construction
$t_{\alpha_k}$ factorizes at least from one of the functions 
${\cal U}_l$, ${\cal F}_l$. The resulting subsector integrals have the 
general form
\begin{eqnarray}\label{EQ:subsec_form}
G_{lk} &=& \int\limits_{0}^{1} d^{N-1}t
\left( \prod\limits_{j=1}^{N-1} t_j^{A_j-B_j\epsilon}  \right)
\frac{{\cal U}_{lk}^{N-(L+1)D/2}}{{\cal F}_{lk}^{N-LD/2}}\, , \quad k=1,\dots ,r
\end{eqnarray}
\end{description}
For each subsector the above steps have to be repeated 
as long as a set ${\cal S}$ 
can be found such that one of the functions 
${\cal U}_{l\dots}$, ${\cal F}_{l\dots}$ vanishes 
if the elements of ${\cal S}$ are set to zero. 
In each subsector new subsectors are created, resulting in a 
tree-like structure after a certain number of iterations. 
The book-keeping can be done with respective multi-indices.
The iteration stops if the functions 
${\cal U}_{l k_1 k_2\dots}$, ${\cal F}_{l k_1 k_2\dots}$
contain a constant term, i.e.
if they are of the following schematic form
\begin{eqnarray}\label{EQ:subsec_UF}
{\cal U}_{l k_1 k_2\dots} &=& 1 +  u(\vec t\,) \\
{\cal F}_{l k_1 k_2\dots} &=& -s_{0} + 
\sum\limits_{\beta} (-s_{\beta}) f_\beta(\vec t\,) \nonumber
\end{eqnarray}
where $u(\vec t\,)$ and $f_\beta(\vec t\,)$ are polynomials in the
variables $t_j$ (without a constant term), and $s_{\beta}$ 
are kinematic invariants, defined through (\ref{DEF:F}).
Thus, after a certain number of iterations, 
each integral $G_l$
is split into a certain number, say $R$, of subsector integrals.
For simplicity we replace the multi-index $k_1 k_2\dots$ stemming 
from the subsector decomposition by a single index which just counts the 
number of generated subsectors.
Now, the produced subsector integrals are exactly of the same form as in 
Eq.~(\ref{EQ:subsec_form}), with the difference that the index 
$k$ now runs from 1 to $R$, the total number of produced subsectors.

Evidently the singular behaviour of the integrand now can be trivially 
read off the exponents $A_j$, $B_j$ for a given subsector integral
($A_j$, $B_j$ are integers). 
The singular behaviour is
manifestly non-overlapping now and thus it is straightforward
to define subtractions. Before doing so a few comments are in order.

The described method cannot always lead to an optimal sector decomposition
in the sense that the integral one starts with is split  
into the {\it minimal} number of subsector integrals, since
obviously even finite integrals would be decomposed if 
a set ${\cal S}$ exists such that ${\cal U}$ or ${\cal F}$ vanish. 
The virtue of the algorithm lies in its easy programmability, as we introduced 
standardized representations suitable for iteration.

For massless 2-loop 4-point functions with 7 propagators and
all external legs on-shell the number of generated subsectors is 
a few hundred. 
For 3-loop 3-point functions with two legs on-shell and 
9 propagators it is a few thousand. Hence the bookkeeping of such 
numbers is only possible on a computer.      

\subsection*{Part III \quad Extraction of the poles}

The third part of the algorithm consists in a pole subtraction procedure for
the subsector integrals $G_{lk}$. As the infrared sensitive
variables now factorize in the subsector integrals, 
one can separate the part of the integrand which leads to $\epsilon$--poles.
 
Explicitly, the following procedure has to be worked through
for each variable $t_{j=1,\dots ,N-1}$ and each
subsector integrand:
\begin{itemize}
\item The integrand of Eq.~(\ref{EQ:subsec_form}), 
characterized by the respective  exponents 
$A_j-B_j\epsilon$ $(j=1,\dots , N-1)$ of $t_j$ 
and the functions of the form (\ref{EQ:subsec_UF}), can for each $t_j$
be written as
\begin{eqnarray}\label{EQsub_step1}
I_j = \int\limits_0^1 dt_j\, t_j^{(A_j-B_j\epsilon)}\, {\cal I}(t_j,\epsilon) 
\end{eqnarray}
If  $A_j\ge 0$, the integration does not lead to an $\epsilon$--pole.
In this case no subtraction is needed and one can go to the next
variable $t_{j+1}$. If  $A_j< 0$, one expands ${\cal I}(t_j,\epsilon)$ into a 
Taylor series around $t_j=0$. Using the definition 
${\cal I}_j^{(p)}(0,\epsilon)=
\partial^p {\cal I}(t_j,\epsilon)/\partial t_j^p\Big|_{t_j=0}$, one obtains
\begin{eqnarray}
{\cal I}(t_j,\epsilon) = \sum\limits_{p=0}^{|A_j|-1}
{\cal I}_j^{(p)}(0,\epsilon)\frac{t_j^p}{p!} + R(t_j,\epsilon) 
\end{eqnarray}
\item
Now  the pole part can be extracted easily, and one obtains
\begin{eqnarray}\label{EQsub_step2}
I_j = \sum\limits_{p=0}^{|A_j|-1} \frac{1}{A_j+p+1-B_j\epsilon}
 \frac{{\cal I}_j^{(p)}(0,\epsilon)}{p!} 
+ \int\limits_{0}^{1} dt_j \, t_j^{A_j-B_j \epsilon} R(t_j,\epsilon) 
\label{tjsubtr}
\end{eqnarray}
By construction the integral containing the remainder term 
$R(t_j,\epsilon)$ does not get poles in $\epsilon$
from the $t_j$-integration anymore. 
For example, in the generic case of a logarithmic divergence, $A_j=-1$, 
$p=0$ and  $R(t_j,\epsilon)={\cal I}(t_j,\epsilon)-{\cal I}_j(0,\epsilon)$.
Since, as long as $j<N-1$, the expression (\ref{tjsubtr}) still contains 
an overall factor $t_{j+1}^{A_{j+1}-B_{j+1}\epsilon}$, it is of the 
same form as (\ref{EQsub_step1}) for $j\to j+1$ and 
the same steps as above can be applied
to it.  
\end{itemize}
After $N-1$ steps
all singular integrations are done analytically and
all poles are extracted. The resulting 
expression can be expanded in $\epsilon$ now. This defines
a Laurent series in $\epsilon$ with coefficients $C_{lk,m}$ 
for each subsector integral $G_{lk}$. 
Since each loop can contribute at most one soft and collinear 
$1/\epsilon^2$ term, the highest possible infrared pole of an $L-$loop 
graph $G$ is $1/\epsilon^{2L}$.
\begin{eqnarray}\label{EQ:eps_series_Glk}
 G_{lk} = \sum\limits_{m=0}^{2L} \frac{C_{lk,m}}{\epsilon^m} + 
{\cal O}(\epsilon) 
\end{eqnarray}
Following the steps outlined above one 
has generated a regular integral representation of the
coefficients $C_{lk,m}$, consisting of  $(N-1-m)$--dimensional finite 
integrals over parameters $t$.

Symmetries of the graph typically lead to equalities between primary
sectors. It is thus useful to calculate the primary sectors $G_l$
\begin{equation}\label{EQ:eps_series_Gl}
 G_{l} = \sum\limits_{k=0}^{R} G_{lk}
\end{equation}
 separately before summing over the $l$ subsectors. 
In this way the symmetry relations 
 provide a nontrivial check of the calculation.

\subsection*{Part IV \quad Calculation of the pole coefficients}

Part four of the algorithm consists in the computation of
the finite subsector integrals. The integrals contributing to the leading pole
give ratios of  polynomials in the Mandelstam variables.
For the coefficient of the subleading pole one generally gets logarithmic
terms. In principle, one can attempt to perform  the
$(N-1-m)$-dimensional  integrations 
of all functions contributing
to the coefficient of the $1/\epsilon^m$ pole analytically. However, 
the sector decomposition produces 
many surface terms which will cancel only after summing up all subsector
integrals, such that the analytical integrations become more and more 
involved for smaller values of $m$, especially if the graphs contain more than
one scale, as for example in the case of 2--loop 4--point functions.
For these functions only the coefficients of the 
leading and subleading poles could be obtained analytically by automatizing 
the integrations using Mathematica~\cite{math}. Pushing the analytical
integrations further is possible only if some of the more complicated 
functions are manipulated "by hand" before feeding them into the
subroutine, but this is tedious in view of the 
large number  of functions to integrate. More powerful
analytical integration routines, specialized to  manipulations
of polylogarithms and Nielsen functions \cite{Nielsen_function} 
would be needed to allow for a complete analytical treatment.

On the other hand, the parametric integral representations
are all very well suited for numerical integration,
as long as the parametric function  ${\cal F}$ has a definite sign.
Then, it contains at most integrable, logarithmic divergences
at the border of the integration domain. These generally present no problems
for the numerical integrators which are on the market.
If ${\cal F}$ is not of a definite sign, which means that one has
to integrate over thresholds, the integrands contain poles
inside the multi-dimensional integration domain. The presence
of these poles typically
considerably slows down, if not hinders at all, the numerical evaluation 
of the integral.  More advanced integration algorithms
like e.g. the one proposed in \cite{jadach} may solve this problem. 
We restrict ourselves to the case of  definite sign here.

In most algebraic programs it is directly possible to create 
FORTRAN functions from a given expression. We fully automatized
the translation of the expressions for the subsector integrals
into the FORTRAN codes.
For every primary sector we calculated the 
integral of the sum of the
subsector integrands (\ref{EQ:eps_series_Gl}) with the
Monte Carlo program BASES \cite{kawabata}.
The integrations of all examples we calculated were totally
stable. The limitation for the numerical integration
comes thus only from CPU time, which increases with the number of 
loops and legs or, correspondingly, with the number of subsector functions.
We will give more detailed information on program parameters in the 
examples below.

\section{Applications}

In this section we apply our method to calculate  several 
nontrivial Feynman diagrams.
To this end we implemented the algorithm outlined in section 2 into
algebraic manipulation programs. To crosscheck the output we
created two independent codes, one  written in Maple~\cite{maple},
the other one in Mathematica~\cite{math}. 
First we show a comparison of the results obtained by our method with 
the results from the analytical calculation
of the  planar~\cite{smirnov} and
non-planar~\cite{tausk} massless double box.  
Then we give results for 
some  2--loop 4--point functions for which analytical results are not yet 
available in the literature, relevant for
the calculation of certain higher order QCD corrections 
(see section \ref{discuss}).
These are the massless planar and non-planar double box 
with one external leg off--shell, given in Figs.~1--3.
As discussed above, our method so far only allows us to calculate
numerical values of these graphs if all the Mandelstam variables 
have the same sign. In any case our result may serve as a
nontrivial check of a future analytical computation
of these graphs. 
Further we will give results for 3--loop
3--point  graphs with two on-shell legs. 

The numerical values given in subsection \ref{dbonsh} contain a relative error 
of one percent. Note that for comparison purposes 
{\em our} conventions for the prefactors ($\Gamma$-functions) should be used
since  multiplying our results
with conversion factors that are a power series in $\epsilon$
may lead to a bad error propagation\footnote{The reason is that 
multiplication with such a conversion factor 
mixes the different pole coefficients in our Laurent series,
which can lead to a larger relative error in the converted result.}.
In subsections 3.2 and 3.3 we use a different prefactor that in 
subsection \ref{dbonsh} because it leads to more compact expressions 
for the analytical result. This conversion may lead to errors in the 
numerical result which are slightly larger than one percent due to the 
error propagation mentioned above. 
  
The numerical 
calculations were done on a DEC--ALPHA workstation running with an EV6
processor.   
The  CPU time needed  ranges from about 5 hours for the
planar two--loop graph up to about 3 days for the 3--loop graph
with 9 propagators. The computer time needed is always dominated by the
finite ${\cal O}(\epsilon^0)$ terms. 

\subsection{Comparison with analytical results}\label{dbonsh}

In this subsection we compare our numerical approach with
the analytical results for massless two-loop box diagrams
calculated only recently~\cite{smirnov,tausk}. 
These results have also been crosschecked analytically meanwhile, 
and the perfect agreement with our results within the error of numerical 
integration confirms the reliability of our method. 

\subsection*{The massless planar double box $B^{P}_{7}(s,t)$}

The graph (see Fig.~\ref{Fig:db_1mass} with leg four on-shell) 
depends on the two kinematical invariants  $s=(p_1+p_2)^2$ and 
$t=(p_2+p_3)^2$. We show a comparison to the analytical result at 
two numerical points, the symmetric point $s=t=-1$ and the asymmetric point 
$(s,t)=(-1,-1/2)$. As stated above, the physical situation $st<0$ 
is not suited for numerical evaluation in a straightforward manner
because thresholds are present.
In the following table we show the comparison between the analytical
result of \cite{smirnov}
and our numerical result for the relevant Laurent
coefficients $c_m^P(s,t)$ 
of the massless planar double box $B^{P}_7(s,t)$, defined through 
Eq.~(\ref{EQ:param_rep}) and the subtraction algorithm:
\begin{eqnarray}
B^{P}_7(s,t)&=&(-1)^7\,\Gamma(3+2\epsilon)
\sum_{m=0}^4\frac{c_m^P(s,t)}{\epsilon^m} + {\cal O}(\epsilon)\label{defdb}
\end{eqnarray}
One can check that the agreement is always better than the demanded 
one percent.
\begin{table}[h!]
\begin{center}
\begin{tabular}{|l|l|l|l|l|l|}
\hline
\multicolumn{6}{|c|}{}\\  
\multicolumn{6}{|c|}{$(s,t)=(-1,-1)$}\\
\multicolumn{6}{|c|}{}\\  
\hline
& $c_4^P$ & $c_3^P$ & $c_2^P$ & $c_1^P$& $c_0^P$  \\ \hline
analytical & -2.    &  6.     & 4.9167 &  -11.495 & -13.801  \\ \hline   
numerical  & -2.0000&  6.0000 & 4.9188 &  -11.492 & -13.811 \\ 
\hline
\hline
\multicolumn{6}{|c|}{}\\   
\multicolumn{6}{|c|}{$(s,t)=(-1,-2)$}\\
\multicolumn{6}{|c|}{}\\  
\hline
& $c_4^P$ & $c_3^P$ & $c_2^P$ & $c_1^P$& $c_0^P$  \\ \hline
analytical &-1.     & 3.8664 & -0.38116 & -9.2384 & -2.9973\\ \hline
numerical  &-1.0000 & 3.8664 & -0.38059 & -9.2377 & -2.9990\\ \hline
\end{tabular}\end{center}
\caption{The massless planar double box. Comparison between analytical
and numerical result for the points $(s,t)= (-1,-1)$ and $(-1,-2)$. 
}
\end{table}

\subsection*{The massless non-planar double box $B^{NP}_{7}(s,t,u)$}

Without using momentum conservation, 
the graph (see Fig.~\ref{Fig:np7_1mass_m1} with leg one on-shell) 
depends on the three kinematical invariants  $s=(p_1+p_2)^2$, 
$t=(p_2+p_3)^2$ and $u=(p_1+p_3)^2$. In the physical region 
one invariant is positive, leading always to an imaginary part. 
To avoid the corresponding threshold
in the numerical integration we treated $s,t,u$ as independent parameters.  
We compared to the analytical result of \cite{tausk}
by calculating numerically the coefficients 
$c_m^{NP}(s,t,u)$ of the Laurent series of $B^{NP}_{7}(s,t,u)$ 
(defined with the same conventions as in  (\ref{defdb})) 
at the symmetric point  $s=t=u=-1$ and the asymmetric
point $(s,t,u)=(-1,-2,-3)$, see Table 2.
\begin{table}[h!]
\begin{center}
\begin{tabular}{|l|l|l|l|l|l|}
\hline
\multicolumn{6}{|c|}{}\\  
\multicolumn{6}{|c|}{$(s,t,u)=(-1,-1,-1)$}\\
\multicolumn{6}{|c|}{}\\  
\hline
& $c_4^{NP}$ & $c_3^{NP}$ & $c_2^{NP}$ & $c_1^{NP}$& $c_0^{NP}$ \\ \hline
analytical & -1.75    &  3.     &22.828  &-113.63  & 395.26 \\ \hline   
numerical  & -1.7500&  2.9960 &22.818  &-113.75  & 393.08 \\ 
\hline
\hline
\multicolumn{6}{|c|}{}\\   
\multicolumn{6}{|c|}{$(s,t,u)=(-1,-2,-3)$}\\
\multicolumn{6}{|c|}{}\\  
\hline
& $c_4^{NP}$ & $c_3^{NP}$ & $c_2^{NP}$ & $c_1^{NP}$& $c_0^{NP}$ \\ \hline
analytical &-0.4167  &0.9310  &5.8586 &-42.760 &162.81 \\ \hline
numerical  &-0.4167  &0.9295  &5.8748 &-42.614 &164.16\\ \hline
\end{tabular}\end{center}
\caption{The massless non-planar double box. Comparison between analytical
and numerical result for the points $(s,t,u)= (-1,-1,-1)$ and $(-1,-2,-3)$.}
\end{table}

\subsection{Two--loop massless 4--point functions, one leg off-shell}

Now we turn to the calculation of graphs where fully analytical results 
do not exist yet. 
The Mandelstam variables $s$, $t$, $u$ are defined as above.
If an external leg $p_k$ is off-shell or on-shell but massive,
we write $p_k^2=m_k^2$. For the 4--point functions under consideration, 
exactly one leg is off-shell. 
Then momentum conservation implies $s+t+u=m_k^2$, but it has to be
emphasized that this constraint has {\em not} been used to obtain the 
analytical results, in order to be able to compare them to numerical
results for unphysical kinematics as well. 

For sums of Feynman parameters we use short-hand notations 
like e.g. $x_i+x_j+x_k+x_l=x_{ijkl}$.

\subsection*{The graph $B^{P}_{7,1mass}(s,t,u,m_4^2)$}
\begin{figure}[ht]
\hspace{4.5cm}
    \epsfxsize = 6cm
    \epsffile{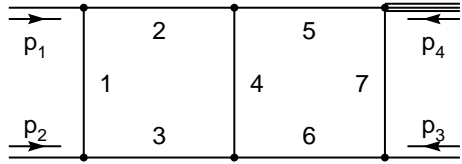}
\caption{\label{Fig:db_1mass}{\em The planar double box with leg 4 off-shell.}}
\end{figure}
With the labeling as in Fig.~\ref{Fig:db_1mass} one finds
for the functions ${\cal U}$, ${\cal F}$:
\begin{eqnarray}
{\cal U} &=& x_{123} x_{567} + x_{4} x_{123567} \nonumber\\
{\cal F} &=& \quad(-s) (x_2 x_3 x_{4567} + x_5 x_6 x_{1234} 
                         + x_2 x_4 x_6 + x_3 x_4 x_5) \nonumber\\
          &&  +(-t) x_1 x_4 x_7  
            + (-m_4^2) x_7 ( x_2 x_4 + x_5 x_{1234} )      
\end{eqnarray}    
The sector decomposition produces
about 200 subsector integrals. For the leading and subleading pole
we get the following analytical result:
\begin{eqnarray}
B^{P}_{7,1mass} &=& \Gamma^2(1+\epsilon)\,(-m_4^2)^{-2\epsilon}\frac{1}{s^2 t}\left(\frac{1}{\epsilon^4} - \frac{2}{\epsilon^3}\left[\,\log(s/m_4^2)+\log(t/m_4^2)\,\right] \right)+ {\cal O}(\frac{1}{\epsilon^2})
\end{eqnarray}
Numerically we find for the points $(-1/3,-1/3,-1/3,-1)$ and $(-1/2,-1/3,-1/6,-1)$:
\begin{eqnarray}
B^{P}_{7,1mass}(-1/3,-1/3,-1/3,-1) =  
\Gamma^2(1+\epsilon)\left(
                              -\frac{27.000}{\epsilon^4} 
                              -\frac{118.65}{\epsilon^3} 
                              -\frac{239.6}{\epsilon^2} 
                              -\frac{305.8}{\epsilon} 
                              -164.1 \right)
\end{eqnarray} 
\begin{eqnarray}
B^{P}_{7,1mass}(-\frac{1}{2},-\frac{1}{3},-\frac{1}{6},-1) =  
\Gamma^2(1+\epsilon)\left(
                              -\frac{12.000}{\epsilon^4} 
                              -\frac{43.005}{\epsilon^3} 
                              -\frac{58.68}{\epsilon^2} 
                              -\frac{20.86}{\epsilon} 
                              + 97.63 \right)
\end{eqnarray} 

\subsection*{The graph $B^{NP}_{7,1mass,a}(s,t,u,m_1^2)$}
\begin{figure}[ht]
\hspace{4.5cm}
    \epsfxsize = 6cm
    \epsffile{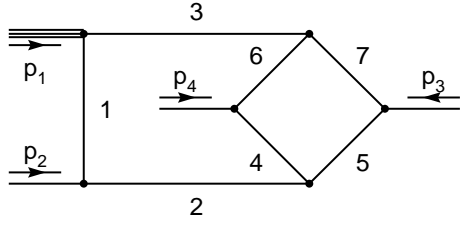}
\caption{\label{Fig:np7_1mass_m1}{\em The non-planar double box with 
leg 1 off-shell.}}
\end{figure}
With the labeling as in Fig.~\ref{Fig:np7_1mass_m1}, the functions 
${\cal U}$, ${\cal F}$ are given by:
\begin{eqnarray}
{\cal U} &=& x_{123} x_{4567} + x_{46} x_{57} \nonumber\\
{\cal F} &=& \quad(-s) (x_2 x_3 x_{4567} + x_2 x_6 x_7 + x_3 x_4 x_5) \nonumber\\
          && +(-t) x_1 x_4 x_7 + (-u) x_1 x_5 x_6
             +(-m_1^2) x_1 ( x_6 x_7 + x_3 x_{4567} )      
\end{eqnarray}      
We note that this graph contains linear IR divergences,
as it is the case if all legs are massless \cite{tausk}. 
In a gauge theory, numerator functions would be present to prevent 
singularities of such strength. 
Since the subtractions are more complicated if linear divergences 
are present, the subsector
integrands are more complicated as well. This leads to larger
FORTRAN functions and thus slows down the numerical computation
in comparison with the similar graph  $B^{NP}_{7,1mass,b}$
below, which has poles stemming only from logarithmic divergences. 
It also requires a refined analytic integration routine. 
The sector decomposition produces
about 250 subsector integrals. For the leading and subleading pole
we obtain the following analytical result:
\begin{eqnarray}
B^{NP}_{7,1mass,a} &=& 
\Gamma^2(1+\epsilon)\,(-m_1^2)^{-2\epsilon}\frac{1}{4s^2 t\,u}
 \Bigl\{\frac{1}{\epsilon^4}\,
 \Bigl[s+t+u-m_1^2 + \frac{t\,u}{m_1^2} + \frac{1}{2}(t+u)\Bigr]\nonumber\\
&& + \frac{1}{\epsilon^3}\,\Bigl[3(s+t+u+m_1^2)\nonumber\\
&&\qquad +\log(s/m_1^2)\,[-2(s+t+u-m_1^2) - 2\frac{t\,u}{m_1^2}-(t+u)]\nonumber\\
&&\qquad +\log(t/m_1^2)\,[-2(s+t+u-m_1^2) + 2\frac{t\,u}{m_1^2}+3t-u] \nonumber\\
&&\qquad +\log(u/m_1^2)\,[-2(s+t+u-m_1^2) + 2\frac{t\,u}{m_1^2} +3u-t]\Bigr]
\Bigr\} \nonumber\\
&&+ {\cal O}(\frac{1}{\epsilon^2})
\end{eqnarray}
Numerically we find for the points $(-1/3,-1/3,-1/3,-1)$ and $(-1/2,-1/3,-1/6,-1)$:
\begin{eqnarray}
B^{NP}_{7,1mass,a}(-\frac{1}{3},-\frac{1}{3},-\frac{1}{3},-1) = 
\Gamma^2(1+\epsilon)\left(
                              -\frac{8.997}{\epsilon^4} 
                              -\frac{101.7}{\epsilon^3} 
                              +\frac{592.7}{\epsilon^2} 
                              +\frac{3340.}{\epsilon} 
                              +18522. \right)
\end{eqnarray} 
\begin{eqnarray}
B^{NP}_{7,1mass,a}(-\frac{1}{2},-\frac{1}{3},-\frac{1}{6},-1) =  
\Gamma^2(1+\epsilon)\left(
                              -\frac{5.504}{\epsilon^4} 
                              -\frac{87.98}{\epsilon^3} 
                              +\frac{296.6}{\epsilon^2} 
                              +\frac{1753.}{\epsilon} 
                              + 11741.\right)
\end{eqnarray} 

\subsection*{The graph $B^{NP}_{7,1mass,b}(s,t,u,m_3^2)$}
\begin{figure}[ht]
\hspace{4.5cm}
    \epsfxsize = 6cm
    \epsffile{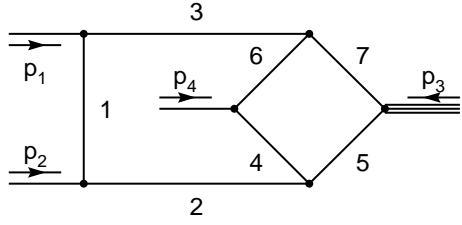}
\caption{\label{Fig:np7_1mass_m3}{\em The non-planar double box with leg 3 
off-shell.}}
\end{figure}
With the labeling as in Fig.~\ref{Fig:np7_1mass_m3} one obtains
for the functions ${\cal U}$, ${\cal F}$:
\begin{eqnarray}
{\cal U} &=& x_{123} x_{4567} + x_{46} x_{57} \nonumber\\
{\cal F} &=& \quad(-s) (x_2 x_3 x_{4567} + x_2 x_6 x_7 + x_3 x_4 x_5) \nonumber\\
          &&  + (-t) x_1 x_4 x_7 + (-u) x_1 x_5 x_6 \nonumber\\
          &&  +(-m_3^2) ( x_5 x_7 x_{12346} + x_2 x_4 x_7 + x_3 x_5 x_6 )      
\end{eqnarray}      
The sector decomposition produces about 230 subsector integrals. 
For the leading and subleading pole
we obtain the following analytical result:
\begin{eqnarray}
B^{NP}_{7,1mass,b} &=& \Gamma^2(1+\epsilon)\,(-m_3^2)^{-1-2\epsilon}\,
\frac{1}{2s\, t\,u}\Bigl\{-\frac{1}{2\epsilon^4}\,(t+u+4m_3^2) \nonumber\\
&& + \frac{1}{\epsilon^3}\Bigl[-(t+u)\log(s/m_3^2)\nonumber\\
&&\qquad +(4m_3^2-t+u)\log(t/m_3^2) \nonumber\\
&&\qquad +(4m_3^2+t-u)\log(u/m_3^2)\Bigr] \Bigr\}+ {\cal O}(\frac{1}{\epsilon^2})
\end{eqnarray}
Numerically we find for the points $(-1/3,-1/3,-1/3,-1)$ and $(-1/2,-1/3,-1/6,-1)$:
\begin{eqnarray}
B^{NP}_{7,1mass,b}(-\frac{1}{3},-\frac{1}{3},-\frac{1}{3},-1) = 
\Gamma^2(1+\epsilon)\left(
                              -\frac{31.50}{\epsilon^4} 
                              -\frac{108.8}{\epsilon^3} 
                              -\frac{2.560}{\epsilon^2} 
                              +\frac{779.8}{\epsilon} 
                              + 2395. \right)
\end{eqnarray} 
\begin{eqnarray}
B^{NP}_{7,1mass,b}(-\frac{1}{2},-\frac{1}{3},-\frac{1}{6},-1) = 
\Gamma^2(1+\epsilon)\left(
                              -\frac{40.50}{\epsilon^4} 
                              -\frac{203.9}{\epsilon^3} 
                              -\frac{357.3}{\epsilon^2} 
                              +\frac{409.9}{\epsilon} 
                              + 4608.\right)
\end{eqnarray} 
We recall that all the numbers given were calculated for
unphysical kinematics (all particles ingoing) in order to have
positive definite denominators.
The numerical results for the 1--mass two-loop graphs 
as given above do not correspond to 
a physical situation. Nevertheless, they can easily be related to
a physical process by the following reasoning.
In the case of a $1\rightarrow 3$
process, e.g. a virtual particle with a squared momentum of $m^2$
decaying into 3 massless particles,
$s,t,u$ and $m^2$ are positive. By factoring out $(-m^2)$ in the respective
functions ${\cal F}$,
one gets again positive definite integrands. Especially, for a one--mass 
2--loop graph with $N$ propagators, one finds 
\begin{eqnarray}\label{EQ1to3}
G^{2-loop}(s,t,u,m^2) &=& (-m^2-i\delta)^{-N+4-2\epsilon }
G^{2-loop}(s/m^2,t/m^2,u/m^2,1)\nonumber\\
                   &=& \frac{1-\theta(m^2)[1-(-1)^N\,e^{2\pi i \epsilon }]}{
     \vert m^2 \vert^{N-4+2\epsilon }} G^{2-loop}(s/m^2,t/m^2,u/m^2,1)
\end{eqnarray} 
We introduced an infinitesimal imaginary part $i\delta$ where
necessary.  This shows that our method allows 
to calculate 2--loop  scalar integrals which appear for example
in the calculation of the NNLO QCD corrections of $e^+e^-\rightarrow 3\, jets$, 
as explained in more detail in section \ref{discuss}.

\subsection{Three--loop massless 3--point functions, one leg off-shell}

Now we want to present results for 3--loop 3--point integrals
with two legs on-shell. These contain only one scale which 
can be factored out of the integrand, such that one has
to calculate a pure number which can be done numerically once and 
forever. We restrict ourselves to only two examples, the graphs
shown in Figs.~\ref{Fig:ms8b} and \ref{Fig:bbt}. 

\subsection*{The graph $MS_8(s)$}
\begin{figure}[ht]
\hspace{4.5cm}
    \epsfxsize = 6cm
    \epsffile{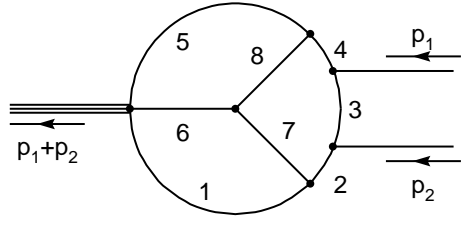}
\caption{\label{Fig:ms8b}{\em A 3--loop Mercedes-Star (MS) topology.}}
\end{figure}
With the labeling as in Fig.~\ref{Fig:ms8b},  
 the functions ${\cal U}$, ${\cal F}$ read:
\begin{eqnarray}
{\cal U} &=& x_{23478} x_5 x_{16} + x_{234} ( x_6 x_8+x_1 x_{68} + x_7 x_{568})
             +x_1 x_7 x_{68} + x_8 (x_1 x_6 + x_5 x_7)  \nonumber \\
{\cal F} &=& (-s) \bigl[  x_1 x_5 ( x_6 x_{23478}+ x_7 x_8)
         + x_1 x_4 (  x_6 x_8 + x_7 x_{568}) \nonumber\\&&\qquad
         + x_2 x_5 ( x_8 x_{167} + x_6 x_7 )
         + x_2 x_4 ( x_{17} x_{58} + x_6 x_{1578}) \bigr]      
\end{eqnarray}      
The sector decomposition produces about 1000 subsector integrals.
We note that because of the symmetries of the graph
only 5 of the 8 primary sectors as defined in Eq.~(\ref{EQ:primary_sectors})
have to be calculated, i.e.
$MS_{8}^{\,\rm{sec}_1}=MS_{8}^{\,\rm{sec}_5}$,  
$MS_{8}^{\,\rm{sec}_2}=MS_{8}^{\,\rm{sec}_4}$, 
$MS_{8}^{\,\rm{sec}_7}=MS_{8}^{\,\rm{sec}_8}$. 
On the other hand, 
the recalculation of these identical sectors are a good check
of the algebraic/numerical computer routines. 
For the leading, subleading  and subsubleading pole
we find the following analytical result:
\begin{eqnarray}
MS_8(s) &=& \frac{\Gamma^3(1+\epsilon)}{(-s-i\delta)^{2+3\epsilon}}\left(
\frac{1}{36\epsilon^6}+\frac{5\zeta(2)}{36\epsilon^4}\right)
+{\cal O}(\frac{1}{\epsilon^3})
\end{eqnarray}
Numerically we find:
\begin{eqnarray}
MS_8(s) = \frac{\Gamma^3(1+\epsilon)}{(-s-i\delta)^{2+3\epsilon}}\left(
                               \frac{0.02778}{\epsilon^6} 
                              +\frac{0.0000}{\epsilon^5} 
                              +\frac{0.2288}{\epsilon^4} 
                              -\frac{0.6692}{\epsilon^3} 
                              -\frac{0.6152}{\epsilon^2} 
                              +\frac{2.005}{\epsilon} 
                              + 17.85 \right)
\end{eqnarray} 

\subsection*{The graph $BBT_9(s)$}
\begin{figure}[ht]
\hspace{4.5cm}
    \epsfxsize = 6cm
    \epsffile{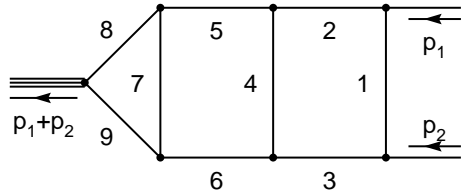}
\caption{\label{Fig:bbt}{\em The 3--loop box-box-triangle (BBT) graph}}
\end{figure}
With the labeling as in Fig.~\ref{Fig:bbt}, 
 the functions ${\cal U}$, ${\cal F}$ are given by:
\begin{eqnarray}
{\cal U} &=& x_{1234} x_7 x_{89} + x_{789} ( x_{123}x_{456}+x_4 x_{56}  )
\nonumber \\
{\cal F} &=& (-s) \bigl[ x_2 x_3 ( x_{4567} x_{89} + x_7 x_{456} ) 
                   +  ( x_3 x_5 + x_2 x_6 ) x_4 x_{789} 
       +  ( x_3 x_8 + x_2 x_9 ) x_4 x_7 \nonumber\\
    && \qquad+  x_5 x_6 x_{1234} x_{789}
       +  ( x_6 x_8 + x_5 x_9 ) x_7 x_{1234}
       +  x_8 x_9 ( x_{1234} x_{567}  + x_{123} x_4 )   \bigr]      
\end{eqnarray}      
The sector decomposition produces about 2400 subsector integrals. 
For the leading and  subleading  pole
we obtain the following analytical result:
\begin{eqnarray}
BBT_9(s) &=&
\frac{\Gamma^3(1+\epsilon)}{(-s-i\delta)^{3+3\epsilon}}
\left(-\frac{1}{36\epsilon^6}
+{\cal O}(\frac{1}{\epsilon^4})\right)\nonumber
\end{eqnarray}
Numerically we find:
\begin{eqnarray}
BBT_9(s) = \frac{\Gamma^3(1+\epsilon)}{(-s-i\delta)^{3+3\epsilon}}\left(
                              -\frac{0.02778}{\epsilon^6} 
                              +\frac{0.0000}{\epsilon^5} 
                              -\frac{0.6852}{\epsilon^4} 
                              -\frac{2.072}{\epsilon^3} 
                              -\frac{6.613}{\epsilon^2} 
                              -\frac{25.07}{\epsilon} 
                              - 40.42 \right)
\end{eqnarray} 

\section{Discussion}\label{discuss}

We want to discuss now the phenomenological applicability
of our method to standard QCD processes. We will focus
on reactions with massless particles in the loop.
The first step of a multi-loop calculation is to express
the corresponding amplitudes, preferably decomposed in colour 
and helicity  space,
in terms of basic tensor and scalar integrals.
In the analytical approach one tries to express
all tensor integrals by a certain set of scalar integrals,
so-called master integrals. In a numerical approach it may
be more convenient to calculate certain tensor integrals
directly. In a Feynman parameter language this amounts
to the calculation of integrals of the type (\ref{EQ:param_rep}),
where one has in addition a polynomial ${\cal A}$ of 
Feynman parameters in the numerator.
The generalization of our method to include these nontrivial 
numerators is straightforward. 
One way to proceed is to carry along the corresponding polynomial 
${\cal A}$ 
through all the steps in addition to ${\cal F}$ and ${\cal U}$, 
iterating the decomposition until ${\cal A}_{lk}$  contains also a constant term. 
This leads to
an expression of type (\ref{EQ:subsec_form}) with the integrand multiplied
by  ${\cal A}_{lk}$. Part III and IV then work in exactly the
same way as above. Since the presence of numerators in general
improves the IR behaviour, the number of necessary
subsector decompositions will be reduced, which typically shortens
the computation time of the diagram under consideration as compared to
the scalar case.

\begin{table}[ht]
\begin{center}
\begin{tabular}{|c|l|l|l|l|}
\hline
process & kinematics & scalar integrals  & analytical & numerical  \\
\hline\hline
&&&&\\
Drell-Yan, DIS&$2\rightarrow 1$ & 3-loop, 3-point, 1 mass & no & yes \\
%\hline
to $N^3LO$ &$2\rightarrow 2$ & 2-loop, 4-point, 1 mass & no & yes $(*)$\\
%\hline
&$2\rightarrow 3$ & 1-loop, 5-point, 1 mass & yes & -- \\
&&&&\\
\hline\hline
&&&&\\
$e^+e^-\rightarrow jjj,jj\gamma,j\gamma\gamma$&$1\rightarrow 3$ & 2-loop, 4-point, 1 mass & no  & yes \\
%\hline
to $NNLO$ &$1\rightarrow 4$ & 1-loop, 5-point, 1 mass &yes& --\\
%\hline
&&&&\\
\hline\hline
&&&&\\
$PP,P\bar P\rightarrow jj,j\gamma,\gamma\gamma$&$2\rightarrow 2$ & 2-loop, 4-point  & yes & yes $(*)$ \\
%\hline
to $NNLO$&$2\rightarrow 3$ & 1-loop, 5-point  & yes$^\ddagger$ & -- \\
%\hline
&&&&\\
\hline
\end{tabular}
\caption{\em Knowledge on scalar integrals needed for various processes. 
In the column ``analytical'', results existing in the literature 
to ${\cal O}(\epsilon^0)$ are listed; those marked with $\ddagger$ 
are known to all orders in $\epsilon$. 
The column ``numerical'' shows where  
our method can improve the situation. 
The asterisk $(*)$ indicates that more powerful numerical integrators 
than the ones used in the present work
would be needed to deal with thresholds.\label{TAB:DY}}
\end{center}
\end{table}

In  Table (3) we list some integrals entering 
the calculation of virtual corrections to
phenomenologically relevant processes. 
We indicate for which ones analytical results exist in the literature
and where our numerical method improves the present situation. 
For the calculation of the NNLO QCD corrections 
of $e^+e^- \rightarrow jjj,jj\gamma,j\gamma\gamma$
two--loop box graphs with one massive external leg 
are needed. As has been demonstrated above these diagrams 
can be calculated numerically with our method. This is also true for
3--loop 3--point functions needed for a $N^3LO$ calculation 
of deep inelastic scattering (DIS) and the
Drell--Yan process\footnote{We note that this refers only to the
calculation of the partonic cross section; for a complete analysis, the
splitting functions also have to be known to this order.}. 
Although our algorithm produces integrable functions which in
principle always allow for a numerical evaluation, the presence
of thresholds inside the integration region worsens  
the convergence properties of an integration routine.
These cases are marked with an asterisk $(*)$ in Table 1. 
It means that more efficient numerical integrators than 
the ones we used would be needed. 

After having calculated the purely virtual corrections
of say a $2\rightarrow N$ process involving $L$--loop 
integrals, the latter  have to be combined  
with the  $2\rightarrow N+1$ corrections which include 
$(L-1)$--loop integrals and where one has to integrate over
the extra (unobserved) particle. This integration produces
soft/collinear $\epsilon$--poles which require
the knowlegde of the respective $(L-1)$--loop integrals 
to ${\cal O}(\epsilon^2)$, at least in the corresponding IR regions, 
since they have to be combined with these singular phase space integrals.
However, most of the existing  
analytical results are known only to ${\cal O}(\epsilon^0)$.
In the context of our method, expansion of the results for the 
virtual integrals up to higher orders in $\epsilon$ 
constitutes no principle problem, only the size of the FORTRAN routines 
will increase. 

\medskip

In conclusion, we have presented a simple, constructive 
subtraction algorithm to deal with IR divergent multi-loop
integrals. Although the iterated sector decomposition produces
in general a relatively large number of subsector integrals, we
demonstrated that by means of automatization 
our method allows for the numerical computation of highly 
nontrivial Feynman diagrams of two-- and three--loop type.
We pointed out that with our numerical approach, 
certain higher order QCD calculations can be tackled
without waiting for further developments of analytical methods.       

\newpage

{\bf\Large Acknowledgments}

\vspace*{5mm}

We thank Bas Tausk for a fruitful exchange of ideas. His analytical 
calculations served not only as a benchmark for our algorithm, 
but also triggered the rigorous development from the initial idea
to the final code. Further we would like to thank V.~A.~Smirnov for 
useful discussions. 
We also are grateful to  J.~Ph.~Guillet for 
helpful discussions and comments on the manuscript. 
G.H. would like to thank the LAPTH for its hospitality 
during various visits. \\
This work was supported by the EU Fourth Training Programme  
''Training and Mobility of Researchers'', Network ''Quantum Chromodynamics
and the Deep Structure of Elementary Particles'',
contract FMRX--CT98--0194 (DG 12 - MIHT).

\end{document}